\documentclass[11pt]{JHEP3}
\usepackage{graphicx}
\usepackage{epsfig}
\usepackage{amssymb}
\input epsf.tex

\newcommand{\intx}{\int d^4 x\,}

\newcommand{\intvecx}{\int d^3 x\,}

\newcommand{\veck}{{\bf k}}

\newcommand{\ep}{\epsilon}

\newcommand{\kp}{\kappa}
\newcommand{\lm}{\lambda}
\newcommand{\rh}{\rho}
\newcommand{\sg}{\sigma}

\newcommand{\ph}{\phi}

\newcommand{\om}{\omega}

\newcommand{\Gm}{\Gamma}

\newcommand{\half}{\frac{1}{2}}

\newcommand{\Tr}{\mbox{Tr}\,}

\newcommand{\phd}{\ph^{\dagger}}

\newcommand{\eela}[1]{\label{#1}\end{equation}}
\newcommand{\eeala}[1]{\label{#1}\end{eqnarray}}
\newcommand{\be}{\begin{eqnarray}}
\newcommand{\ee}{\end{eqnarray}}
\newcommand{\dcp}{\delta_{\rm cp}}

\newcommand{\NCS}{N_{\rm cs}}
\newcommand{\NW}{N_{\rm w}}

\newcommand{\mueff}{\mu_{\rm eff}}
\title{Simulations of Cold Electroweak Baryogenesis: Finite time quenches}

\author{Anders Tranberg$^{a,b}$, Jan Smit$^{c}$ and Mark Hindmarsh$^{a}$\\
$^{a}$) Department of Physics and Astronomy, University of Sussex,\\
Brighton, East Sussex BN1 9QH, United Kingdom.\\
$^{b}$) DAMTP, University of Cambridge \\
Wilberforce Road, Cambridge, CB3 0WA, United Kingdom.\\
$^{c}$) Institute for Theoretical Physics, University of Amsterdam, \\
Valckenierstraat 65, 1018 XE Amsterdam, the Netherlands.\\
}


\preprint{ITFA-2006-36, DAMTP-2006-88}

\abstract{The electroweak symmetry breaking transition may
  supply the appropriate out-of-equilibrium conditions for
  baryogenesis if it is triggered sufficiently fast. 
This can happen at the end of low-scale inflation, prompting baryogenesis to occur during tachyonic preheating
of the Universe, when the potential energy of the inflaton
  is transfered into Standard Model particles. With the proper amount
  of CP-violation present, the observed baryon number asymmetry can be
  reproduced. Within this framework of Cold Electroweak Baryogenesis,
  we study the dependence of the generated baryon asymmetry on the
  speed of the quenching transition. We find that there is a separation
  between ``fast'' and ``slow'' quenches, 
which can be used to put
  bounds on the allowed Higgs-inflaton coupling. We also clarify the 
  strong Higgs mass dependence of the asymmetry reported in a
  companion paper  \cite{Tranberg:2006ip}.}

\begin{document}

% SECTION: INTRODUCTION

\section{Introduction\label{introduction}}

The possibility of generating the observed baryon asymmetry at the
electroweak phase transition dates back just over twenty years
\cite{Kuzmin:1985mm}. The Standard Model provides baryon number violating
processes as well as CP-violation and potentially departure from
thermal equilibrium. The task at hand is to discover a scenario in
which these ingredients come together and generate the correct
amplitude of the baryon asymmetry.

Cold Electroweak Baryogenesis was proposed as an alternative to
Electroweak Baryogenesis at a bubble wall in a first order phase
transition 
\cite{Garcia-Bellido:1999sv,Krauss:1999ng,Copeland:2001qw}. 
In the Minimal Standard Model such a transition is ruled out by bounds on the
Higgs mass 
\cite{Csikor:1998eu,Kajantie:1996kf,pdg}, 
and Cold Electroweak
Baryogenesis relies on an out-of-equilibrium symmetry breaking
transition triggered by the evolution of an inflaton field. 

Inflation 
is supported by measurements of the CMB \cite{Spergel:2006hy} 
and although the
precise origin of the accelerated expansion has not been confirmed,
models based on a scalar field rolling in a suitable potential can
successfully reproduce observational signals. The energy scale of
inflation is constrained by observations, but is 
ususally
guided by expectations of physics beyond the Standard Model. 
In Cold Electroweak Baryogenesis, we apply a minimal extension
to the Standard Model, a single inflaton, and assume that the
energy scale of inflation is the Electroweak
scale. Although this does not discard the possibility of further physics at
higher energies, it allows for a self-contained description of
baryogenesis. This because a period of electroweak-scale inflation will reduce
the impact from whatever came before. 

In section 2 we will introduce the relevant details of Cold Electroweak Baryogenesis. Tachyonic preheating with varying quench times
is studied in section 3, with no CP-violation, 
where we observe the behaviour of CP-even observables. These results
are then used in section 4, where we add CP-violation and compare the full simulation results to linear treatments based on the CP-even observables,
as in \cite{Tranberg:2006ip}. 
We 
interpret
the quench-time and mass dependence 
in terms of the winding 
around zeros of the Higgs field. 
Our conclusions are
in section 5.

% SECTION:  ELEMENTS OF COLD ELECTROWEAK BARYOGENESIS

\section{Cold Electroweak Baryogenesis\label{CEB}}

A number of conditions are to be met for Cold Electroweak Baryogenesis to
be 
successful:
\begin{itemize}
\item 
A period of inflation has to take place that ends at sufficiently low energy, such
  that the reheating temperature is well below the electroweak
  scale. 
Then baryon number changing processes (sphaleron transitions) cannot wash out a
  previously generated asymmetry. Such low scale inflation can be
made to agree with observational constraints, with some allowance for
parameter tuning \cite{German:2001tz,Copeland:2001qw,vanTent:2004rc}. 
\item 
Baryon number violation has 
to be effective while the
  system is out of equilibrium. 
The finite-temperature transition 
is a crossover rather than a phase transition 
in the Minimal Standard Model, 
\cite{Csikor:1998eu,Kajantie:1996kf,pdg}, 
and hence the system stays close to equilibrium. 
The phase transition can however be first-order in 
supersymmetric models with an enlarged Higgs sector (see for instance \cite{Laine:1998qk}).

We pursue here a different route.
Out-of-equilibrium
  conditions can be introduced by triggering electroweak symmetry
  breaking through a coupling of the Higgs field to a rolling inflaton
  at or after the end of inflation. This results in a tachyonic
  instabiliy, which in turn 
can result
in change of baryon number 
\cite{Garcia-Bellido:1999sv,Krauss:1999ng,Copeland:2001qw,Garcia-Bellido:2003wd,Tranberg:2003gi,Tranberg:2006ip}. 
\item Although the gauge fields come out of equilibrium, an asymmetry
  is only created in the presence of CP-violation, the strength of
  which will determine the magnitude of the baryon asymmetry. The quark sector of the
  Minimal Standard Model seems unlikely to provide the required
  CP-violation \cite{Shaposhnikov:1987tw,Shaposhnikov:1988pf,Gavela:1994ds,Gavela:1994dt,Smit:2004kh}.  The lepton sector 
has not been studied in similar detail. Here, as in \cite{Tranberg:2003gi,Tranberg:2006ip}, we use only a simple CP-violating interaction composed of Higgs and gauge fields and consider it as representing the generic case. 
\end{itemize}

A simple implementation of the scenario is obtained by adding
effective CP violation to the SU(2)-Higgs sector of the SM, with
action
(we use the metric $(-+++)$)
\be
\label{action}
S=
-\int d^{3}{\bf x}\,dt \bigg[\frac{1}{2g^{2}}\Tr F^{\mu\nu}F_{\mu\nu}+(D^{\mu}\phi)^{\dagger}D_{\mu}\phi
+ \ep
+\mu_{\rm eff}^{2}
\phi^{\dagger}\phi+\lambda(\phi^{\dagger}\phi)^{2}+\kappa\,\phi^{\dagger}\phi\,
\Tr F^{\mu\nu}\tilde{F}_{\mu\nu}\bigg]\nonumber,
\ee
The effective mass $\mueff$ is supposed to depend on time, e.g.\
due to the coupling to an inflaton field $\sg$, 
\be
\mu_{\rm eff}^{2}=\mu^{2}-\lambda_{\sigma\phi}\sigma^{2},
\label{massterm}
\ee
with a potential to appropriately generate low-scale inflation. 
In `inverted hybrid inflation' models \cite{Lyth:1996kt,Copeland:2001qw,vanTent:2004rc},
$\sg$ is very small during inflation as it rolls away from the origin. 
After inflation has ended it becomes substantial such
that $\mueff^2$ flips sign when $\sg^2=\mu^2/\lambda_{\sigma\phi}$. 
Much later $\sg$ and $\ph$ settle near their vacuum values and the eigenmodes 
and eigenvalues of the inflaton-Higgs mixing mass matrix determine the 
masses and other properties of the ensuing spinless particles
\cite{vanTent:2004rc}.

The speed of the transition may be characterized by
\be
u=\left.\frac{\sqrt{2}}{m_H^3}\,\frac{d\mu_{\rm eff}^{2}}{dt}\right|_{\mu_{\rm eff}^{2}=0}.
\label{u}
\ee
where we used the Higgs mass to set the scale.
In \cite{vanTent:2004rc} we 
assumed
that viable baryogenesis would require
a sufficiently fast quench $|u|\gtrsim 0.15$.
For very fast quenches, the system is expected to be very much
out-of-equilibrium, and in the limit of infinitely slow quenches, the
system should stay in equilibrium throughout the symmetry breaking
transition.  One issue is to identify a transition between ``fast''
and ``slow'' transitions. 
This will allow us to constrain $\lambda_{\sigma\phi}$ and/or the shape of the inflaton potential around the time of symmetry breaking. Note that in the particular implementation of \cite{vanTent:2004rc}, inflation ends long before electroweak symmetry breaking, so that there is some freedom to tune the inflaton speed to accommodate a fast quench.

Setting aside the inflaton and the shape of its potential, 
we model the effective mass parameter by a linear form
\be
\label{linmassterm}
\mu_{\rm eff}^{2}(t)=&\mu^{2}\left(1-\frac{2t}{t_{Q}}\right),&0<t<t_{Q},\\
\mu_{\rm eff}^{2}(t)=&-\mu^{2},&t>t_{Q}.
\ee
with $t_{Q}$ introduced as the quench time. 
In this model the
Higgs v.e.v.\ and mass are given by the usual formulas
$v^2 = \mu^2/\lm$ and $m_H^2 = 2\mu^2$, and $\ep = \mu^4/4\lm$ is chosen
to set the energy density of the vacuum to zero. Furthermore
\be
|u|=\frac{1}{\mu t_{Q}} = \frac{\sqrt{2}}{m_H t_Q}.
\ee
For $|u|=0.15$, we find 
$m_{H}t_{Q}\approx 9$.

The baryon asymmetry now depends on three parameters
\begin{itemize}
\item The quench time $t_Q$.
\item The strength of CP-violation, encoded in the coefficient of the
  CP-violating term, 
which we write in terms of a dimensionless
  $\delta_{\rm cp}$,
\be
\kappa=\frac{3\,\delta{\rm cp}}{16\pi^{2}m_{W}^{2}}.
\label{introdlcp}
\ee
In \cite{Tranberg:2006ip} we found that the dependence on $\delta_{\rm cp}$ is
  linear at least for $\delta_{\rm cp}<1$. 
For example, at $m_{H}=2m_{W}$ this leads to
\be
\frac{n_{B}}{n_{\gamma}}=-(0.32\pm0.04)\times 10^{-4}\,\dcp.
\ee
\item The mass of the Higgs 
boson, 
written as
\be
\left(\frac{m_{H}}{m_{W}}\right)^{2}=\frac{8\lambda}{g^{2}}.
\ee
In 
\cite{Tranberg:2006ip}, the mass dependence was seen to be quite complicated
in the case of 
an instantaneous quench. We will confirm here, that the mass dependence is 
also present with non-instantaneous quenches; even the sign of
the baryon asymmetry depends on $m_H$.
\end{itemize}

Shortly after $\mueff^2$ has turned negative,
the occupation numbers of the fields grow (faster than)
exponentially with time. Initially, the non-linear terms in the 
equations of motion are not very large and it is reasonable to estimate 
$\langle \phd\ph \rangle$ using a Gaussian approximation. The equation
\be
\ddot\ph - \nabla^2\ph + \mueff^2 \ph =0,
\ee
where
\be
\mueff^2 = -M^3(t-t_c),
\quad t_c =t_Q/2,
\quad M =(\mu t_Q/2)^{-1/3} \mu,
\ee
can be solved analytically, 
assuming the initial state at $t=0$ to be the free-field vacuum
(see e.g.\ \cite{Garcia-Bellido:2002aj}, for the instantaneous quench see \cite{Smit:2002yg}). 
Following the steps taken in \cite{Garcia-Bellido:2002aj}, the (Fourier transform of the)
equal-time two-point function is found to be given by
\be
\langle \ph_\veck\phd_\veck \rangle&=& |f_k(t)|^2,
\\
f_k(t) &=& C_{1k} Bi(\tau-k^2/M^2) + C_{2k} Ai(\tau-k^2/M^2),
\quad \tau = M(t-t_c),
\\
C_{1k} &=& -\frac{\pi}{\sqrt{2\om_k}}
[Ai'(-\om_k^2/M^2) + i\om_k Ai(-\om_k^2/M^2)],
\quad
\om_k = \sqrt{\mu^2 + k^2},
\\
C_{2k} &=& \frac{\pi}{\sqrt{2\om_k}}
[Bi'(-\om_k^2/M^2) + i\om_k Bi(-\om_k^2/M^2)].
\ee
Since the Airy functions behave for
large $z$ like $Ai(z)\propto e^{(-2/3)z^3}$, $Bi(z) \propto e^{(2/3)z^3}$,
it can be shown that
occupation numbers of the unstable modes with $k\lesssim M\sqrt{\tau}$ 
grow very rapidly once $\tau > 0$, and a classical approximation can be
made for $\tau \gtrsim 2$ \cite{Garcia-Bellido:2002aj}. In fact, since quantum and classical
evolution are identical in form, the classical evolution
may already be started at time zero ($\tau = -M t_c$), which is what we
do in the numerical simulations (this is the analog of the ``just the half''
initial conditions in the case of the instantaneous quench \cite{Smit:2002yg}). 

An important point to make is that energy is not conserved during a mass quench of the type (\ref{linmassterm}),
because $\mueff$ depends explicitly on time. Differentiating the Hamiltonian
corresponding to the action (\ref{action})
and using the equations of motion one finds
\be
\frac{d H}{d t} = \intvecx \frac{d\mueff^2}{d t}\, \phd\ph.
\ee
Hence during the linear quench there is a change in energy density
\be
\Delta \rh (t) = -\frac{2\mu^2}{t_Q} \int_0^{t} dt'\, \langle \phd\ph \rangle
(t')
\approx
-\frac{2\mu^2}{t_Q}\, 4 \int_{0}^{t} dt' d^{3}{\bf k} |f_k(t')|^2,
\label{energy_loss}
\ee
where we made the Gaussian approximation in the last step (4 is the number
of real components of the Higgs doublet).
Evaluation of this analytic expression shows that the depletion in energy can
be substantial\footnote{The divergence of the integral over momenta is
taken care of by renormalization, which may be 
approximated
by including only the
unstable modes $k < M\sqrt{\tau}$. 
The upper boundary of the integration over time
has to be before
nonlinearities become important, 
which is typically before the end of the quench at 
$t=t_Q$, except for a very fast quench.
}.

In the case when the mass
term comes from a coupling to an inflaton field (\ref{massterm}), 
the energy
extracted from the gauge-Higgs system is 
transferred into inflaton energy.
Had we included the dynamics of the inflaton itself, this energy would not be lost but would come back through additional reheating as the inflaton oscillates and eventually decays. Including a dynamical inflaton does however complicate the system somewhat, since 
it implies unknown parameters such as $\lm_{\sg\ph}$ and other masses and
couplings of a potential $V(\sg,\ph)$.

We shall use the approximation (\ref{linmassterm}) in the present work to bypass the complications of the additional degree of freedom, because it is easy to implement and solve analytically at early times, 
and because it can represent a source of mass quenching other than a coupling to an inflaton. We are therefore not bound to a specific model. 

When calculating the final asymmetry, we will need to know the final reheating temperature. We will then specialise to the case of a Higgs-inflaton coupling, 
assume $m_\sg > m_W$ and 
that all the {\it initial} energy is equipartitioned between the SM particles
at a temperature $T_{\rm reh} < m_W$.

% SECTION: CP-EVEN OBSERVABLES

\section{No CP violation\label{CPeven}}
In order to understand the underlying gauge and Higgs dynamics, it is
useful to study the evolution of various 
observables in absence of CP violation. 
The
CP-violation can then, at least at early times, be thought of as a small perturbation on this background.

The numerical implementation is identical to the one introduced in
\cite{Tranberg:2003gi,Tranberg:2006ip}. In short, the action (\ref{action}) is
discretised on a lattice, and the classical equations of motion solved
in real time. This is done for a CP-symmetric ensemble of initial
conditions, reproducing the Higgs field correlators of the vacuum
state at $t=0$, 
when the Higgs potential is $V=\ep + \mu^2\phd\ph$
in the Gaussian approximation.
The initial gauge field is set by
$A^{\mu}=0$ 
and we impose the Gauss constraint on the gauge momenta 
for a given random Higgs field background. Observables are averaged over the ensemble. 

The observables of interest 
(in a periodic volume $L^3$) are the Higgs field averaged over
the volume,
\be
\overline{ \phi^{2}} = \frac{1}{L^3}\int d^{3}x\, \phd\ph
 = \frac{1}{L^3}\int d^{3}x\, \frac{1}{2}\Tr\Phi^{\dagger}\Phi,
\quad
\Phi=(i\tau_2\ph^*,\phi),
\label{higgssq}
\ee
the volume-averaged magnetic field,
\be
\overline{ B^2} =\frac{1}{L^3}\int d^{3}x\, \Tr F_{ij}F_{ij},
\label{Bsq}
\ee
the distribution of Higgs winding number,
\be
N_{\rm w}
= \frac{1}{24\pi^{2}}\int d^{3}x\epsilon_{ijk}\Tr
U^{\dagger}\left(\partial_{i}U\right)U^{\dagger}\left(\partial_{j}U\right)U^{\dagger}\left(\partial_{k}U\right),
\quad
U=\frac{\Phi}{\sqrt{\half\Tr \Phi^{\dagger}\Phi}},
\ee
the width of the Chern-Simons number distribution 
\be
\Delta_{\rm cs}(t)=\langle [N_{\rm cs}(t)-N_{\rm cs}(0)]^2\rangle, 
\quad
N_{\rm cs}(t) -N_{\rm cs}(0) =\int_{0}^{t} dt \int d^{3}x\,
\frac{1}{16\pi^{2}}\Tr F^{\mu\nu}\tilde{F}_{\mu\nu},
\label{Ncssq}
\ee
and its time derivative, the Chern-Simons number diffusion rate or
susceptibility,
\be
\Gamma=\frac{d\Delta(t)}{dt}.
\label{Gamma}
\ee
In equilibrium, $\Gamma$ is the sphaleron rate.
 
The dynamics of a tachyonic electroweak phase transition has been
studied in some detail in terms of these and related observables in
\cite{Garcia-Bellido:2003wd,Diaz-Gil:2005qp} 
(with an inflaton), 
in the limit of an instantaneous quench in 
\cite{Tranberg:2003gi,vanderMeulen:2005sp,Tranberg:2006ip}, 
and in terms of suitably defined particle numbers in \cite{Skullerud:2003ki}.
The 
above
observables are all global quantities (integrals over space).
Since $A^{\mu}=0$ initially, also $N_{\rm cs}(0)=0$\footnote{We recall
that $N_{\rm cs} =- \intvecx \ep_{jkl}\Tr A_j\left(F_{kl} + i\frac{2}{3}A_k A_l\right)/16\pi^2$.}. 
The Chern-Simons number and the winding number 
can be changed by an integer through a ``large'' 
gauge transformation, 
but $N_{\rm cs}-N_{\rm w}$ is gauge invariant. 
Classical vacuum configurations have $N_{\rm cs}-N_{\rm w}=0$, and are gauge equivalent to $N_{\rm cs}=N_{\rm w}=0$.

We now motivate the study of one other observable, the distribution of
$\phd\ph$ over the volume, in particular its magnitude near $\phd\ph=0$.
The study in \cite{vanderMeulen:2005sp} concentrated on 
{\em local} observables, such as Chern-Simons and winding-number densities,
in an attempt to clarify how CP violation causes an asymmetry in
the final Chern-Simons number. It was observed that the transition 
produced initially many centers with high winding-number density, 
dubbed `half-knots' since their winding number in small balls is roughly
$\pm 1/2$. They can only disappear or be created when the Higgs length 
$\sqrt{\phd\ph}$ becomes zero in their center.
This happens at early times because of the rapid growth of long distance 
modes, and then the number of half-knots rapidly diminishes,
but when the volume-averaged Higgs length has grown substantially,
$\overline{\ph^2}= \mathcal{O}(v^2)$, such zeros have become rare. 
Subsequently the Higgs field `overshoots' the minimum of its potential, 
rolls back and moves again towards zero, and when $\overline{\ph^2}$ goes 
through its first minimum, 
new zeros in the Higgs length occur, enabling the creation of new
half-knots. In this way second, third, \ldots, generation half-knots
were observed \cite{vanderMeulen:2005sp} near the minima of the 
oscillating $\overline{\ph^2}$. Since the gauge field is initially very
small,
CP violation is ineffective 
in influencing
the creation or annihilation of the first generation half-knots. 
However, by the time of the first minimum of $\overline{\ph^2}$, the gauge 
field has grown substantially, CP violation can be effective and
may change the balance and
cause an asymmetry in $\langle N_{\rm cs}\rangle$.      
Because of the importance of Higgs zeros, we shall in section 
\ref{higgszeros} 
present results for
the distribution of the local Higgs field length,
a simple histogram of
$\phd(x)\ph(x)$ for various quench times.

We found in \cite{Tranberg:2006ip} that the asymmetry depends
strongly
on the Higgs mass. The extreme cases seem to be
$m_{H}=2m_{W}\simeq 161$ GeV and $m_{H}=\sqrt{2}\, m_{W}\simeq 114$ GeV
\footnote{In 1+1 dimensions the mass dependence 
is complicated \cite{Smit:2002yg}, which could
also be the case here.}. In addition to the quench time dependence, we are interested in how this mass dependence comes about. Below, we will show plots for these two Higgs masses in parallel, to demonstrate the differences throughout. Since the physical Higgs masses, at least in the Minimal Standard Model are constrained to be within $114$ and $200$ GeV 
\cite{pdg}, both options are (marginally) allowed.

% SUBSECTION: HIGGS FIELD

\subsection{Higgs field\label{Higgs}}

\FIGURE{
\epsfig{file=./pictures/phisq_vtau.eps,width=7cm,clip}
\epsfig{file=./pictures/phisq_vtau_s2.eps,width=7cm,clip}
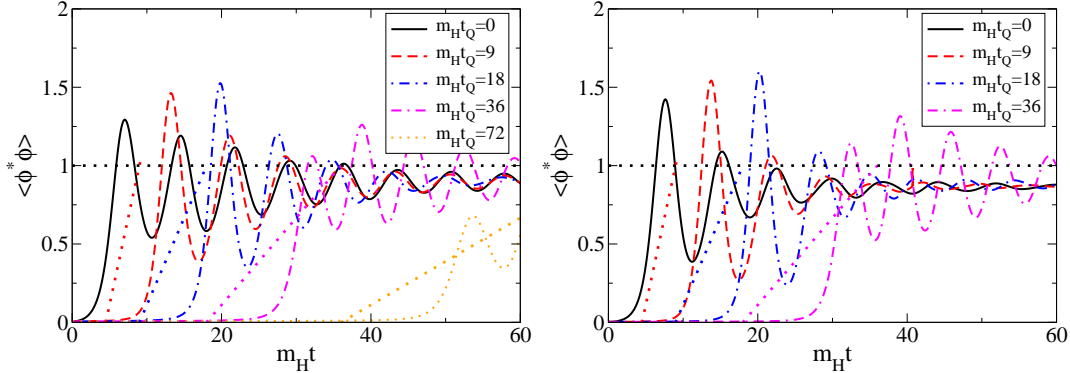
\caption{The 
normalized
Higgs expectation value 
$2\langle\overline{\ph^2}\rangle/v^2$ 
vs
time, for different quench
  times (full lines). Dotted lines show 
$v^{2}(t)/v^2$ 
correspondingly
  colour coded. 
Left: $m_{H}=2m_{W}$, right: $m_{H}=\sqrt{2}m_{W}$.
}
\label{Higgs_tdepv}
}

As the Higgs mass parameter changes from positive to negative, the
naive
Higgs 
expectation value goes from $0$ for
$t<t_{Q}/2$, to $v^{2}(t)= -\mu_{\rm eff}^{2}(t)/\lambda$ up until $t=t_{Q}$, after which
 it stays at $v^{2}= \mu^{2}/\lambda$. 
 
Figure \ref{Higgs_tdepv} shows the evolution of the 
normalized
Higgs average value
$\langle 2\overline{\ph^2}\rangle/v^2$
(cf.\ (\ref{higgssq})) 
for various quench times. Dotted lines show 
$v^{2}(t)/v^2$
for the different $t_{Q}$. For small quench times, the
symmetry breaking time is determined by the time it takes for the dynamics
to perform the rolling off
the local maximum of the potential at $\ph=0$. 
For $m_{H}t_{Q}<36$, $v^{2}(t)$ reaches $v^2$ before 
$\langle 2\overline{\phi^{2}}(t)\rangle$ reaches $v^{2}(t)$. For larger quench times, $\langle 2\overline{\phi^{2}}(t)\rangle$ catches up with $v^{2}(t)$, 
and oscillates around it before settling around $v^2$.
In the limit of infinite quench time one would expect it to follow $v^2(t)$
closely, except for finite-temperature corrections.

We notice that the amplitudes of the first maximum and minimum 
are
quench-time dependent, and also mass dependent. Although the qualitative behaviour is the same, $m_{H}=\sqrt{2}m_{W}$ leads to 
a lower first Higgs minimum.

The evolution of the Higgs field determines the energy loss 
in (the first half of) Eq.~(\ref{energy_loss}) 
due to the time-dependent effective mass.
We integrate the actual
numerical $\langle\phi^{2}\rangle$ up to time $t_{Q}$ to find that for $t_{Q}=(0,9,13.5,18,36,72)$
Eq.~(\ref{energy_loss}) predicts 
\be
|\Delta \rho/\rho_{\rm initial}|\simeq~0,~0.06,~0.10,~0.19,~0.66,~0.84.
\ee
By directly calculating the energy, we find 
\be
|\Delta \rho/\rho_{\rm initial}|\simeq~0,~0.04,~0.08,~0.17,~0.67,~0.83.
\ee 
Note that for the very
slowest quenches, more than half the energy is lost. For these slow
quenches, at $t=t_{Q}$ the field has already started its oscillation,
and so 
the Gaussian approximation
(the second half of Eq.~(\ref{energy_loss})) does not apply. Integrating
the actual field evolution reproduces the energy depletion.
 
% SUBSECTION: MAGNETIC FIELD

\subsection{Magnetic field\label{Magnetic}}

\FIGURE{
\epsfig{file=./pictures/B2_vs_tau.eps,width=7cm,clip}
\epsfig{file=./pictures/B2_vs_tau_s2.eps,width=7cm,clip}
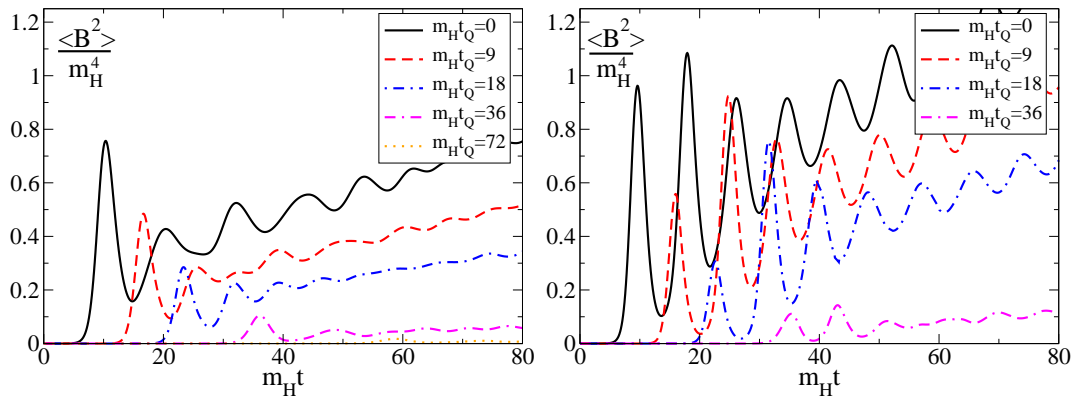
\caption{The evolution of the magnetic field 
$\overline{B^2}$
for different
  $t_{Q}$. $m_{H}=2\,m_{W}$ (left), $m_{H}=\sqrt{2}\,m_{W}$ (right).}
\label{Bfield}
}

As the Higgs field goes through its transition, energy is transfered
to the gauge fields, the occupation numbers of which grow
exponentially \cite{Skullerud:2003ki}. The gauge fields 
acquire energy very fast at first, and then slowly towards what will
be an equipartitioned and thermalised final state, 
figure \ref{Bfield}.
This later transfer is  
quench-time dependent, 
faster quenches lead to a faster transfer of energy.
To some extent, this decrease in energy transfer is due to the 
depletion of energy resulting from the time-dependent 
$\mueff$, 
and the {\it relative} growth $\dot{B^{2}}/B^{2}$ is in fact
similar for different quench times.
The dependence on the Higgs mass is also interesting.
For $m_{H}=\sqrt{2}\, m_{W}$ the transfer seems to be 
driven by the Higgs field oscillations, with the gauge field also oscillating  even for rather late times. In particular the second maximum is much larger than for $m_{H}=2\,m_{W}$.

% SUBSECTION: CHERN-SIMONS DIFFUSION

\subsection{Chern-Simons diffusion\label{CSdiffusion}}

\FIGURE{
\epsfig{file=./pictures/ncssq_vs_tau.eps,width=7cm,clip}
\epsfig{file=./pictures/ncssq_vs_tau_s2.eps,width=7cm,clip}
\caption{
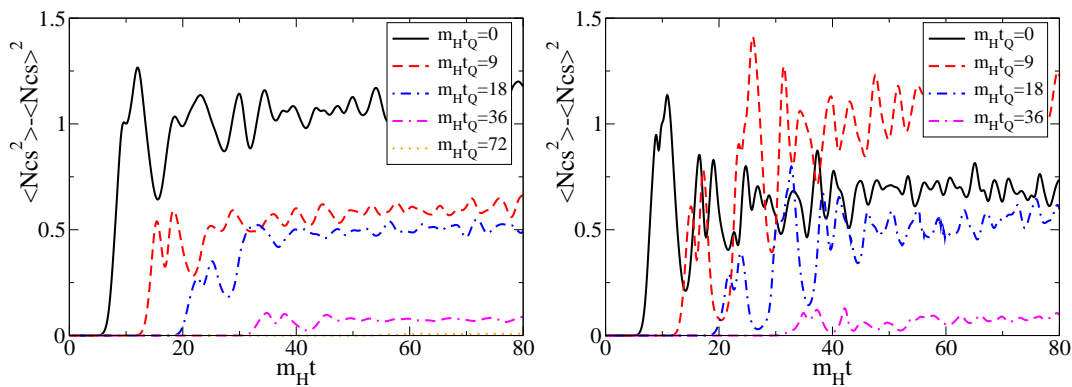
$\langle N_{\rm cs}^{2}\rangle$ for different quench times. $m_{H}=2\,m_{W}$ (left), $m_{H}=\sqrt{2}\,m_{W}$ (right).}
\label{ncssq}
}

In the absence of CP-violation, the ensemble average of the
Chern-Simons number itself is zero, Chern-Simons number being
CP-odd. In our setup, the ensemble is strictly CP-even, so $\langle
N_{\rm cs}\rangle$ is also strictly zero.

We can also calculate the evolution of the width of the
$N_{\rm cs}$
distribution, which will grow as a result of the preheating of the gauge fields, 
similar to
the $B$-field. But also because 
of fluctuations
at finite temperature (and
out of equilibrium, at finite energy density) there is a non-zero
diffusion rate of Chern-Simons number, 
$\Gm(t)=  d\Delta_{\rm cs}/dt$ 
(cf. (\ref{Ncssq})). 
The rate $\Gm(t)$
enters in an
estimate of the baryon asymmetry in section
\ref{withCP}.

Figure \ref{ncssq} shows the evolution of $\Delta_{\rm cs}$ for 
various quench times. 
There is a rapid growth, which in some cases appears to be further
driven by the Higgs field oscillations.
Eventually
$\Delta_{\rm cs}$ settles, 
in accordance with the fact that at the final emerging temperatures 
the equilibrium sphaleron rate is negligible. This is
one of the central features of Cold Electroweak Baryogenesis, 
the
generated asymmetry does not get further diluted by sphaleron transitions.
Similar to the magnetic
field, at least for $m_{H}=\sqrt{2}\,m_{W}$,
the Chern-Simons number seems to be driven by the Higgs
oscillations. However, whereas the growth of the magnetic field has a monotonic dependence on the quench rate, the diffusion rate appears to have a more complicated dependence.

% SUBSECTION: ADDING CP-VIOLATION

\section{Adding CP-violation\label{withCP}}

For small enough values of the CP-violation parameter $\delta_{\rm
  cp}$, it was seen in \cite{Tranberg:2006ip} that the baryon asymmetry is
  linear.
The value of $\dcp$ required to reproduce the observed asymmetry, is comfortably within this linear range. We shall keep
  $\delta_{\rm cp}=1$ throughout, the upper end of the range studied
  in \cite{Tranberg:2006ip}, in order to maximise the generated numerical signal.

Because the CP-violation is small, we have the option to treat it as a
perturbation on the CP-even background described in the previous section. Although we will include the CP-violating term completely
in the dynamics below, we will first apply the early time
approximations introduced in 
\cite{Khlebnikov:1988sr,Garcia-Bellido:1999sv,Garcia-Bellido:2003wd,Tranberg:2003gi} for the case of finite-time quenches.

% INITIAL RISE: LINEAR THEORY

\subsection{Initial rise\label{initialrise}}

\FIGURE{
\epsfig{file=./pictures/init_rise_t0_v2.eps,width=7cm,clip}
\epsfig{file=./pictures/init_rise_t36_v2.eps,width=7cm,clip}
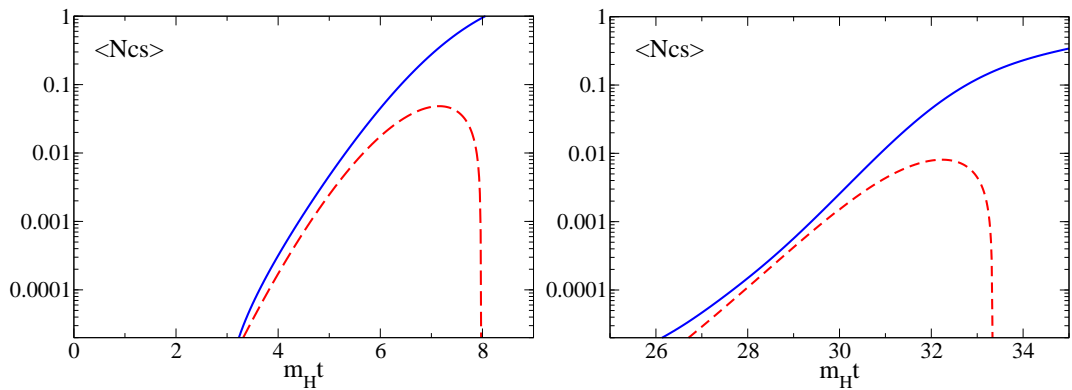
\caption{ The initial rise, comparing the simulation (dashed, red) to
  Eq.~(\ref{linhom}) (full, black), for $m_{H}t_{Q}=0$ (left) and $36$ (right). Notice the logarithmic axis.}
\label{init_rise}
}
For very early times, only very long wavelength modes have large
occupation numbers, both in the Higgs and gauge fields,
since modes only gradually become unstable, starting with the zero mode at time $t_{Q}/2$. 
Treating the CP-violation as a perturbation to the CP-even evolution,
and making a homogeneous approximation, we find for the average Chern-Simons
number
\cite{Smit:2002yg}
\be
\langle N_{\rm cs}\rangle=
\frac{\sqrt{2}\,\dcp (Lm_{H})^{3}}{64\pi^4(1+c)^2}
\,
\frac{\langle\overline{B^{2}}\rangle}{m_{H}^{4}}
\,\frac{\langle\overline{\ph^2}\rangle}{v^2/2}.
\label{linhom}
\ee
The 
values of $B^{2}(t)$ and $\phi^{2}(t)$ are taken from the
simulations, 
section \ref{CPeven}. The constant
$c$ is extracted from the growth of the gauge field 
$\overline{B^{2}}(t)\propto e^{2ct}$.
The linearisation 
assumes exponential growth of the Higgs and gauge fields.
This
description will have to break down at some fairly early time, when the
back-reaction of gauge and Higgs non-linear self-interaction becomes important. Figure
\ref{init_rise} shows the linear approximation compared to the full
simulation for short and long quench times. The agreement is good during the
initial exponential growth, but 
breaks down after about 5 and 10 units 
of $m_H t$ after $\mueff^2$ has gone negative at $t=t_Q/2$,
respectively for $t_Q=0$ and 36.

% EARLY BACK-REACTION: ASYMMETRIC DIFFUSION

\subsection{
Thermodynamic treatment\label{Boltzmann}}

\FIGURE{
\epsfig{file=./pictures/linresallt_v2.eps,width=7cm,clip}
\epsfig{file=./pictures/linresallt_s2.eps,width=7cm,clip}
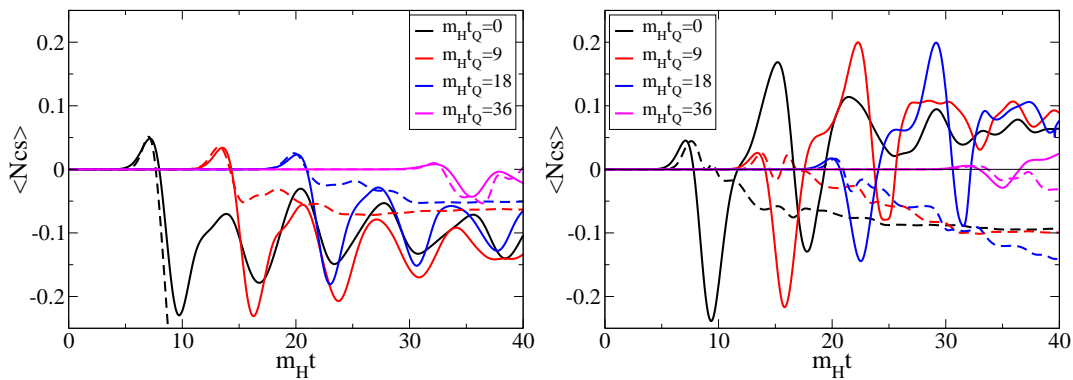
\caption{The average Chern-Simons number in the full simulations (full
  line) and from the 
thermodynamic 
treatment (dashed line). $m_{H}=2\,m_{W}$ (left), $\sqrt{2}\,m_{W}$ (right).}
\label{boltzmann}
}
Beyond the linear approximation, we can apply 
methods from non-equilibrium thermodynamics \cite{Khlebnikov:1988sr,Garcia-Bellido:1999sv,Garcia-Bellido:2003wd,Tranberg:2003gi} 
to estimate the asymmetry. 

One can interpret the CP-violating term as a chemical potential for Chern-Simons number\footnote{Notice that one treats $\phd\ph$ as an space-independent chemical potential.} (cf.\ (\ref{action},\ref{introdlcp})):
\be
\intx \kp \phi^{\dagger}\phi\Tr F\tilde F \leftrightarrow -\int dt\,  \mu_{\rm ch} \NCS,
\quad
\mu_{\rm ch}(t)=\frac{3\delta_{\rm cp}}{m_W^2} 
\frac{d}{dt}\langle \overline{\phi^{2}}(t)\rangle.
\ee
Using the CP-even evolution of the diffusion rate Eq.~(\ref{Gamma}) and the Higgs average Eq.~(\ref{higgssq}),
the average Chern-Simons number can then be estimated through
\be
\label{boltzeq}
\langle N_{\rm cs}\rangle(t)=\frac{1}{T_{\rm eff}}\int_{0}^{t} dt'\,\Gamma(t')\mu_{\rm ch}(t'),
\ee
where $T_{\rm eff}$ was interpreted in
\cite{Garcia-Bellido:1999sv} as the effective temperature of the tachyonic modes. We will not 
elaborate here on such an interpretation, but merely observe that
$T_{\rm eff}$ turns out to decrease roughly linearly with $t_{Q}$, and
that $m_{H}=\sqrt{2}\,m_{W}$ gives much larger values, figure ~\ref{Teff}.

\FIGURE{
\epsfig{file=./pictures/Teff.eps,width=8cm,clip}
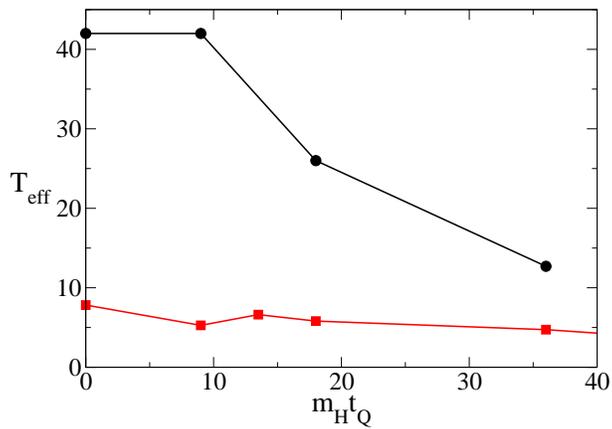
\caption{The effective temperature 
%\JS%\JS%\JS 
in units of $m_H$
as extracted from the thermodynamical treatment. Squares: $m_{H}=2\,m_{W}$, circles: $m_{H}=\sqrt{2}\,m_{W}$.}
\label{Teff}
}

Figure \ref{boltzmann} compares the result of Eq.~(\ref{boltzeq}) to the full
simulation. $T_{\rm eff}$ is chosen to fit the first maximum of the
full simulation. The approximation nicely reproduces the change of
sign of the asymmetry produced by the back-reaction. At later times,
the approximation again breaks down. We will see that this
is precisely the time when the Higgs field acquires a net winding
number \cite{Tranberg:2006ip}, the dynamics of which can apparently not be described by a simple chemical potential
with constant $T_{\rm eff}$. The effective temperatures as a function of $t_Q$ are shown in figure \ref{Teff}.

Notice 
in figure \ref{boltzmann}
that the sign of the asymmetry at later times $m_H t\sim 40$ has
changed again to positive (the sign of $\dcp$) in the case of mass ratio $\sqrt{2}$, which is not captured by the thermodynamic treatment. In principle
the latter might do better, since the oscillations in $\mu_{\rm ch}(t)$ and
$\Gm(t)$ are correlated. In any case, 
replacing the diffusion rate by its time average
\cite{Garcia-Bellido:2003wd}
\be
\int_{0}^{t_{max}}dt'\Gamma(t')\mu_{\rm ch}(t')~~\rightarrow\bar{\Gamma}\int_{0}^{t_{max}}dt'\mu_{\rm ch}(t')
= \frac{3\dcp\bar{\Gamma}v^{2}}{2m_W^2},
\ee
gives a sign of the asymmetry that is definitely equal to that of $\dcp$, 
which may be wrong.

% SUBSECTION FULL SIMULATION

\subsection{Full simulation\label{fullsim}}

\FIGURE{
\epsfig{file=./pictures/plotncs_in_time_all_tau_v2.eps,width=7cm,clip}
\epsfig{file=./pictures/plotncs_in_time_all_tau_s2_v2.eps,width=7cm,clip}
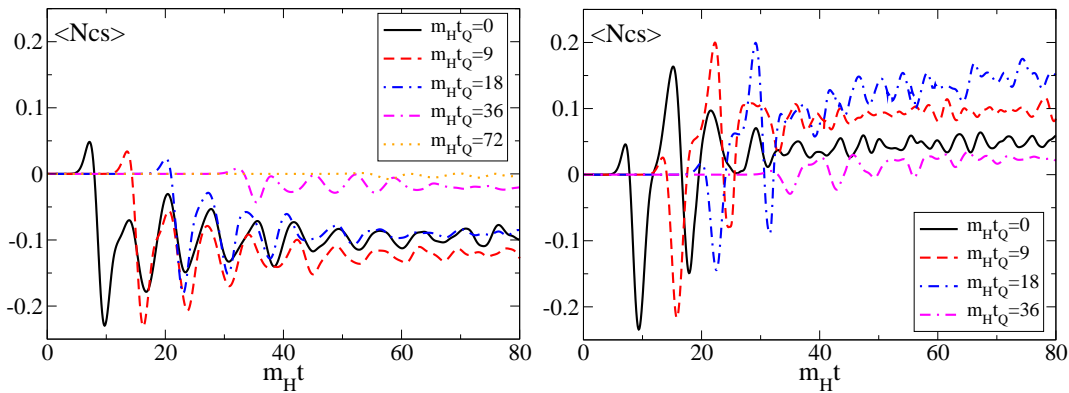
\caption{The evolution of Chern-Simons number in time, $m_{H}=2\,m_{W}$ (left), $m_{H}=\sqrt{2}\,m_{W}$ (right).}
\label{ncsintime}
}

\FIGURE{
\epsfig{file=./pictures/plotnw_in_time_all_tau_v2.eps,width=7cm,clip}
\epsfig{file=./pictures/plotnw_in_time_all_tau_s2_v2.eps,width=7cm,clip}
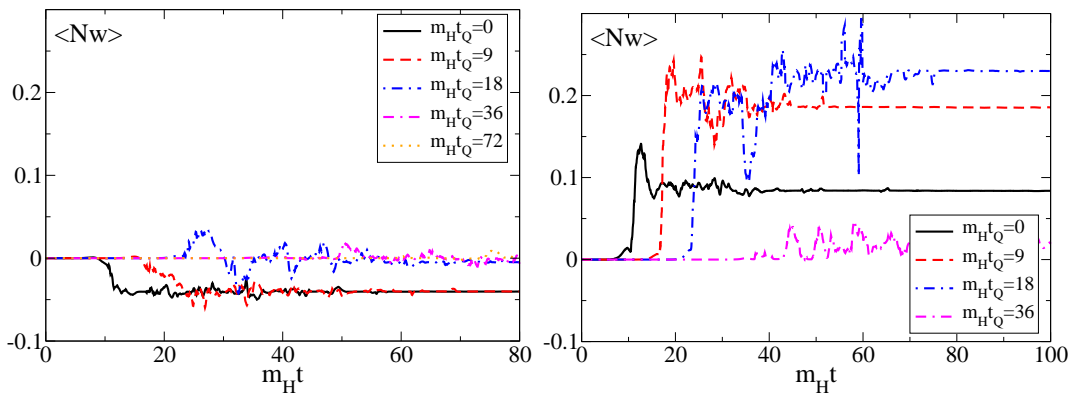
\caption{The evolution of winding number in time, $m_{H}=2\,m_{W}$ (left), $m_{H}=\sqrt{2}\,m_{W}$ (right).}
\label{nwintime}
}

In order to capture the full dependence on quench time and Higgs mass,
we need to include the CP-violation completely in the dynamics. Figure
\ref{ncsintime} shows the 
average
Chern-Simons number for various quench
times. Figure \ref{nwintime} is the corresponding winding number. We 
notice that the mass dependence found in \cite{Tranberg:2006ip} is robust, and not a pathology of an instantaneous quench. For $m_{H}=2\,m_{W}$, the fastest quenches $m_{H}t_{Q}=0,9$ 
lead to an asymmetry of {\em opposite sign} to $\dcp$.
For slower quenches, 
the noise dominates and we can only conclude that the final asymmetriy is
consistent with zero. In contrast, for $m_{H}=\sqrt{2}\,m_{W}$, the asymmetry 
has the {\em same sign} as $\dcp$, and 
it is
maximal for intermediate quench times $m_{H}t_{Q}=18$.
Recall also the maximal boosting of  $\Delta_{\rm cs}$
in section \ref{CPeven} was seen at these quench times. 
In both cases, for $m_{H}t_{Q}=36$ and larger the asymmetry 
appears to vanish.
This is all compiled in figure \ref{final_asym} which shows the final asymmetry 
versus quench time. 

\FIGURE{
\epsfig{file=./pictures/resdivts24.eps,width=10cm,clip}
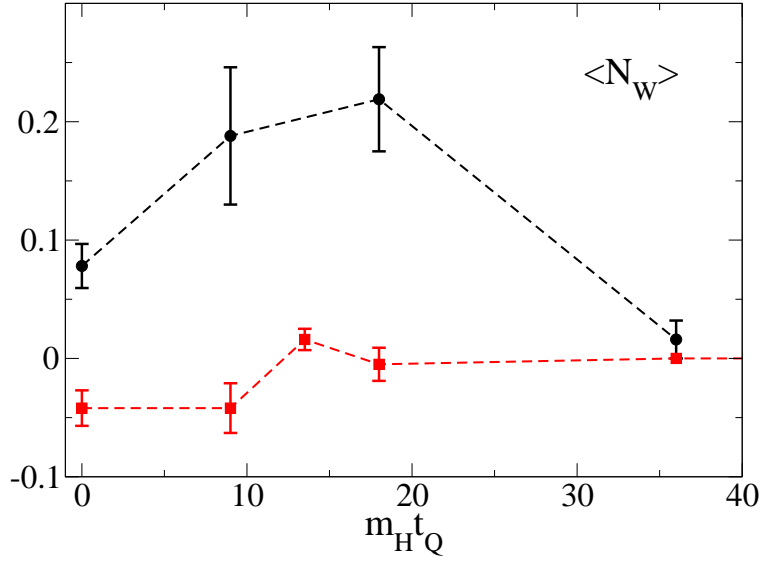
\caption{Final asymmetry vs. quench time for $m_{H}=2\,m_{W}$ (squares) and $\sqrt{2}\, m_W$ (circles).}
\label{final_asym}
}

% SUBSECTION MASS DEPENDENCE

\FIGURE{
\epsfig{file=./pictures/zoom_int9.eps,width=10cm,clip}
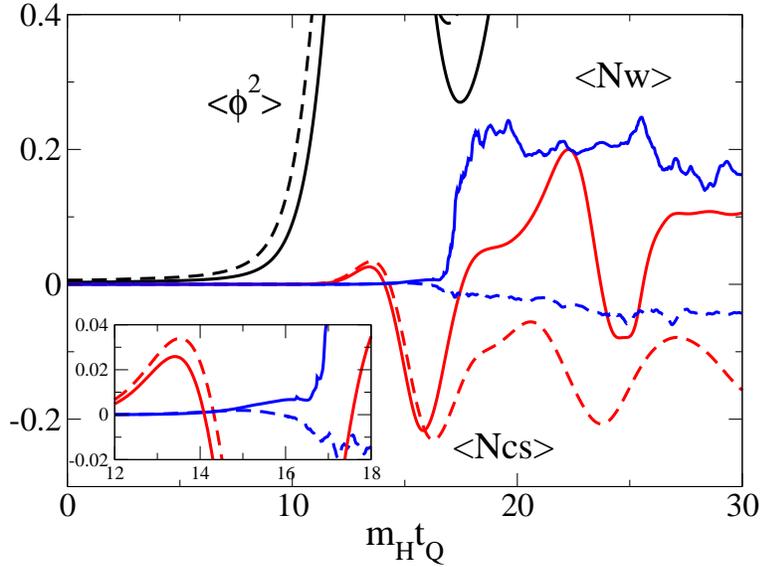
\caption{A close-up of the early evolution of 
$\langle\overline{\ph^2}\rangle$ 
(black), $\langle N_{\rm cs}\rangle$ (red) and $\langle N_{W}\rangle$ (blue). 
Full lines are $m_{H}=\sqrt{2}\,m_{W}$, dashed $m_{H}=2\,m_{W}$.
The quench time $m_{H}t_Q=9$. The inset is a further amplification around
the initial winding number bump.}
\label{zoom_mdep}
}

% SUBSECTION: HIGGS FIELD ZEROS

\subsection{Higgs field zeros\label{higgszeros}}

We have observed in earlier work 
\cite{Tranberg:2003gi,vanderMeulen:2005sp} that the final asymmetry in
$\langle\NCS\rangle$ can already be seen at earlier times in 
$\langle\NW\rangle$, which may be expected from the fact that the 
temperature after the transition is low enough that sphaleron 
transitions are suppressed and the robustness of winding number under 
relatively small changes in the fields. The asymmetry
in $\langle\NW\rangle$ is induced by the CP violating terms in the 
equations of motion, which are very small during the first stages of
the instability, as monitored roughly by $\overline{\ph^2}$ and 
$\overline{B^2}$. 
Somewhat later the asymmetry becomes visible in the 
initial rise and bouncing back of $\langle\NCS\rangle$, and a little later
also in $\langle\NW\rangle$. The rise in $\langle\NW\rangle$ is much 
smaller than in $\langle\NCS\rangle$, presumably since $\NW$ can only 
change when there are zeros in the Higgs field, which are exceptional by 
that time. Still somewhat later, around the time 
$\langle \overline{\phi^{2}}\rangle$ has its first minimum,
$\langle\NW\rangle$ has grown substantially, as
new (second generation) zeros appear in the Higgs field.

This is illustrated in figure \ref{zoom_mdep}, which shows the
evolution of $\langle N_{\rm cs}\rangle$, $\langle N_{\rm w}\rangle$ and 
$\langle \overline{\phi^{2}}\rangle$ at early times. At the time of
the initial bump of $\langle N_{\rm cs}\rangle$ the winding asymmetry 
$\langle N_{\rm w}\rangle$ is still not visible, but a growth of a `bump' 
is discernible by the time $\langle N_{\rm cs}\rangle$ has crossed zero
and reached a negative maximum. When  
$\langle \overline{\phi^{2}}\rangle$ reaches its first minimum 
the asymmetry in $\langle N_{\rm w}\rangle$ grows much faster, which we
interpret as being caused by the asymmetric 
creation of second generation winding centers, 
made possible by zeros in the Higgs lengths.

\FIGURE{
\epsfig{file=./pictures/s4_zeros.eps,width=12cm,clip}
\caption{Histogram of $\phi^{2}(x)$ over the lattice for
  $m_{H}=2\,m_{W}$. Colours correspond to quench times $m_{H}t_{Q}=0$
  (black), $9$ (red), $18$ (green) and $36$ (blue). The four graphs
  correspond to the first
four
minima 
of the Higgs oscillation in each case.
\label{zeross4}
}}

\FIGURE{
\epsfig{file=./pictures/s2_zeros.eps,width=12cm,clip}
\caption{Histogram of $\phi^{2}(x)$ over the lattice for
  $m_{H}=\sqrt{2}\,m_{W}$. Colours correspond to quench times $m_{H}t_{Q}=0$
  (black), $9$ (red), $18$ (green) and $36$ (blue). The four graphs
  correspond to the first four minima of the Higgs oscillation in each case.}
\label{zeross2}
}

To illustrate 
the presence or absence of zeros in the Higgs field
we show 
histograms of $\phi^{2}(x)$ over the
lattice for the first 4 minima of the Higgs oscillation,
in figures \ref{zeross4} and \ref{zeross2}. Figure
\ref{zeross4} is for $m_{H}=2\,m_{W}$, and we see that although the
average is away from zero (figure \ref{Higgs_tdepv}),
there is still a
tail stretching to zero, at least for the first two minima. These will
provide nucleation points for winding. In figure \ref{Higgs_tdepv}, we
also saw that for the smaller Higgs mass $m_{H}=\sqrt{2}\,m_{W}$, the
Higgs minima were somewhat lower. 
It is remarkable,
however how different the distributions in the first minimum look (figure \ref{zeross2}). The
proximity of the distribution bulk to zero (and the fact that
$\phi^{2}(x)\ge 0$) 
results in points aggregating close to zero. 
Qualitatively,
the density of zeros follows the same behaviour as a function of
quench time as the final asymmetry of figure \ref{final_asym}.

%SECTION: KIBBLE MECHANISM

\subsection{Kibble mechanism\label{Kibble}}

\FIGURE{
\epsfig{file=./pictures/wind_hist.eps,width=10cm,clip}
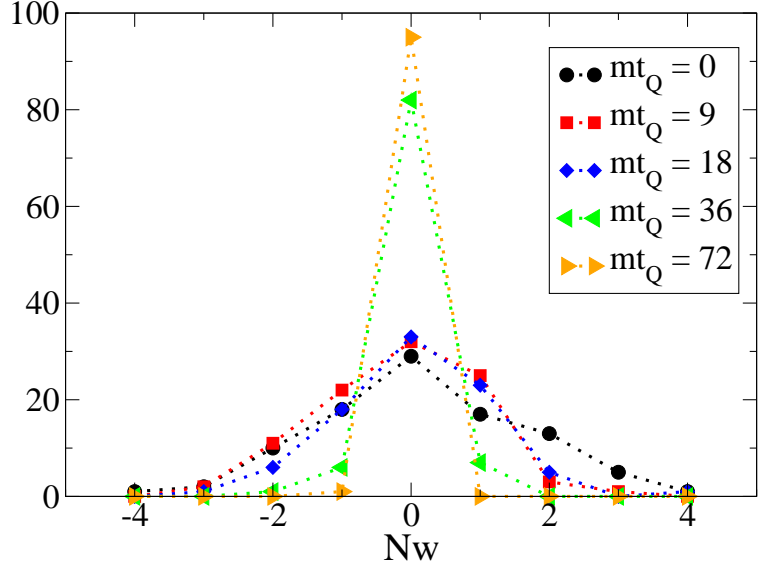
\caption{The final distribution of winding number for various $t_{Q}$. $m_{H}=2\,m_{W}$.}
\label{Kibbleplot}
}

During a symmetry breaking transition, a net density of defects will form through the Kibble mechanism \cite{Kibble:1976sj}. In the present case of 
$O(4)$ symmetry in 3+1
dimensions, these textures have integer 
winding number
in the Higgs field, 
spread out over 
space. The density of gauged
defects can be predicted in terms of the evolution of the correlation
length of the system \cite{Hindmarsh:2000kd}. Numerical
studies often consider thermal quenches with overdamped dynamics
through the transition. In our case, we have an underdamped system
with no 
explicit
coupling to a thermal bath
(higher-momentum modes do play the role of a bath). 
Still, we here illustrate
``fast'' and ``slow'' quenches in terms of the number of defects generated. 

Figure \ref{Kibbleplot} shows the distribution of final 
winding number
over the ensemble. There is a qualitative difference between
$m_{H}t_{Q}= 0$, 9, 18 and $m_{H}t_{Q}= 36$, 72.
Also from this distribution we see that 
$m_H t_Q=18$ still belongs to the regime of `fast' quenches,
whereas $m_H t_Q=36$ is definitely in the `slow' regime.

%SECTION: CONCLUSION

\section{Conclusions\label{conclusions}}

We have 
confirmed 
in a more realistic setting
that including CP-violation in the SU(2)-Higgs
equations leads to baryogenesis, when going through a tachyonic
electroweak transition. The dependence on quench time is significant.
Although our scan of quench time is not fine enough to give a precise
characterisation of `fast' and `slow' quenches, we can say that 
$m_{H}t_{Q}\lesssim 20$ belongs to the first category, corresponding to
\be
|u| = \left|\frac{d\mu_{\rm eff}^{2}(t)}{2\mu^3 d t}\right|_{t=t_{Q}/2}\gtrsim 0.07.
\ee
This constrains a possible underlying Hybrid inflation model in terms of the
Higgs-inflaton coupling $\lambda_{\sigma\phi}$, since
$\mu_{\rm eff}^{2}(t)=\mu^{2}-\lambda_{\sigma\phi}\sigma^{2}$.

It is satisfying that the dramatic mass dependence seen in
\cite{Tranberg:2006ip} is not a result of the instantaneous quench. It is,
however, surprising that for $m_{H}=\sqrt{2}\,m_{W}$ the maximum
asymmetry is not generated at the fastest quenches but for
intermediate $t_{Q}$. 
Both the mass dependence and the quench time dependence can be put down to a coincidence of phases and frequencies of the Higgs and gauge oscillations. This is 
similar to the 
much simpler model in 1+1 dimensions studied in 
\cite{Smit:2002yg}.

An important aspect of the transition is the occurrence of zeros in the Higgs field. CP-violation generates asymmetries in the Chern-Simons number density, which prompts the winding number density to move along as well. The full winding number change is however only realised once the second generation Higgs zeros appear. Although some settling of winding and Chern-Simons number can occur at later generations, we find that the asymmetry is established at the first Higgs minimum.

The
quench time and the mass (through the Higgs self-coupling) influence
the number of zeros, and so determine the 
magnitude of the asymmetry
by allowing more
half-knots to flip their winding number. The detailed dynamics are
very complicated.

The expansion of the Universe is negligible at
electroweak-scale temperatures, but reheating may proceed differently
in the presence of all the other Standard Model fields and in
particular an oscillating inflaton \cite{Diaz-Gil:2005qp}. Still, because the final
asymmetry is largely determined during the first couple of Higgs
oscillations, we expect that the results presented here are reasonably
close to the complete result.

The maximal asymmetry 
occurred in our simulation 
for $m_{H}=\sqrt{2}\,m_{W}$ and 
$m_{H}t_{Q}=18$. 
We can estimate the photon density by distributing the initial energy density over the Standard Model degrees of freedom\footnote{Taking into account the inflaton and details of the lepton sector will not significantly alter the estimate.}, 
which gives:
\be
\frac{n_{B}}{n_{\gamma}}=(0.20\pm0.04)\times 10^{-3}\,\dcp.
\ee
about three times\footnote{The result of \cite{Tranberg:2006ip} is
  based on larger statistics and a fit over a range of $\dcp$. The
  result at zero quench time in figure \ref{final_asym} is about three
  times smaller than the one at $m_{H}t_{Q}=18$.} the result at zero
quench time \cite{Tranberg:2006ip}. This means that we require
$\dcp=3\times 10^{-6}$ or larger to reproduce the observed baryon
asymmetry.

% SECTION: ACKNOWLEDGMENTS

\subsection*{Acknowledgments}
We thank Meindert van der Meulen and Denes Sexty for collaboration on
related subjects and useful comments. A.T. is supported by PPARC Special Programme Grant {\it``Classical Lattice Field Theory''}. Part of this work was conducted on the COSMOS supercomputer, funded by HEFCE, PPARC and SGI, and on the SARA PC-cluster LISA. This work received support from FOM/NWO.

\bibliographystyle{JHEP}
\bibliography{anderslit2JS}

 \end{document}